\documentstyle[12pt]{article}
\author{Serge Galam\footnotemark[1] and Alain Mauger\\
Acoustique et Optique de la Mati\`{e}re Condens\'{e}e\footnotemark[2]\\
Tour 13 - Case 86, 4 place Jussieu, 75252 Paris Cedex 05, France\\[1ex].}
\title{Universal Formulae for Percolation Thresholds  }
\addtocounter{footnote}{0}
\footnotetext{Permanent adress: Groupe de Physique des Solides, T23, Universit\'{e}s Paris 7 et 6,
2 Place Jussieu, 75751 Paris Cedex 05}
\addtocounter{footnote}{1}
\footnotetext{Laboratoire associ\'{e} au CNRS (URA n$^{\circ}$ 800) et
\`{a} l'Universit\'{e}
P. et M. Curie - Paris 6}
\addtocounter{footnote}{1}
\date{Phys. Rev. E \underline{53}, 2177 (1996)}
\begin{document}
\maketitle
\baselineskip 3.3ex
\footskip 5ex
\parindent 2.5em
\abovedisplayskip 5ex
\belowdisplayskip 5ex
\abovedisplayshortskip 3ex
\belowdisplayshortskip 5ex
\textfloatsep 7ex
\intextsep 7ex
\begin{center}
{\em PA Classification Numbers:\/} 64.60 A, 64.60 C, 64.70 P\\
\end{center}

%%%%%%%%%%%%%%%%%%%%%%%%%%%%%%%%%%%%%%%%%%%%%%%%%%%%%%%%%%%%%%%%%%%%%%%
\begin{abstract}

A power law is postulated for both site and bond percolation thresholds.
The formula writes
$p_c=p_0[(d-1)(q-1)]^{-a}d^{\ b}$, where $d$ is the space dimension and $q$
the coordination number. All thresholds up to $d\rightarrow \infty$ are
found to
belong to only three universality classes. For first two classes
$b=0$ for site dilution while $b=a$ for bond dilution. The last one
associated to
high dimensions is characterized by $b=2a-1$ for both sites and bonds.
Classes are defined by a set of value
for $\{p_0; \ a\}$.
Deviations from available numerical
estimates at $d \leq 7$ are within $\pm 0.008$ and $\pm 0.0004$ for high
dimensional
hypercubic expansions at $d \geq 8$.
The formula is found to be also valid for Ising critical temperatures.
\end{abstract}
%%%%%%%%%%%%%%%%%%
\newpage

Percolation theory has been known for several decades [1, 2]. It deals with
the effects of
random dilution of either sites or bonds in a lattice. Flory introduced it
first in the
framework of chemical industry [3]. Upon site or bond dilution
a sharp change is
found to occur in the connectivity of the system at some threshold $p_c$ in
the density
of occupied site or bond. Although it is a purely
geometrical phenomenon, this change can be
described in terms of usual second order phase transitions.
This mapping to critical phenomena
made percolation a full part of the theoretical framework of collective
phenomena and
statistical physics.
It is indeed a very powerful and general tool. Percolation has thus been
applied to
numerous problems
in a large variety of fields, even outside physics (for reviews, see [4, 5,
6, 7]).

However, despite both its success
and its
mathematical ground, percolation theory has resisted exact calculations.
Most known data are
numerical estimates, from both Monte Carlo simulations and series expansions [6, 8]. 
In particular, analytic calculations of percolation
thresholds has
proven to be a rather
difficult task. For instance, twenty years
or so were necessary to prove
the numerical estimate of $p_c=1/2$ for the square bond percolation
threshold [9].
The bond threshold is also known exactly for the two-dimensional honeycomb
and triangular
lattices [9].
The situation is even worse in the case of site percolation.
Indeed, thresholds are known exactly only in the cases of two-dimensional
triangular and Kagom\'{e} lattices [9].
At dimension higher than two, no thresholds were determined exactly.
From simulations, percolation thresholds are found to depend on both the space
dimension $d$ and
the coordination number $q$. Only the
pathological Cayley tree has been solved exactly to yield
$p_c=1/(q-1)$ [10]. This expression which holds for
 both bonds
and sites, yields good results at high dimensions.

Along the general form of the Bethe expression, various empirical
formulas have been tried to yield large classes of thresholds without
success. For instance
the formula $p_c=d/[(d-1)(q-1)]$ was proposed twelve years ago for
site percolation [11].
Another one, $p_c=d/[(d-1)q]$  was suggested  for bond dilution [12].
Unfortunately, both formulas are not satisfactory for all lattices,
especially at low dimensions.
More recently the expression
$p_c=1/\sqrt {q-1}$ was derived for site percolation thresholds
[13, 14]. Very good results are obtained at two dimensions but not at higher dimensions.
Another similar form was also noticed elsewhere [15].
In this paper, for the first time, one unique power law is found to yield
within an excellent
accuracy both site and bond
percolation thresholds  for all lattices at all dimensions. The power law  is
$p_c=p_0\{(d-1)(q-1)\}^{-a}d^{\ b}$.
From a log-log plot all available data ($d\leq 7$) are found to fit on two
straight
lines. In both cases $b=0$ for site dilution
and $b=a$ for bond dilution.
One line includes two-dimensional triangle, square
and honeycomb lattices, which constitute the first class, characterized by
$\{p_0=0.8889; \ a=0.3601\}$ for
site dilution and by
$\{p_0=0.6558; \ a=0.6897\}$ for bond dilution. Two-dimensional Kagom\'{e}
and all
 other regular lattices (for
$d\geq 3$) align on the other unique line and constitute the second class,
 characterized by
$\{p_0=1.2868; \ a=0.6160\}$, and
$\{p_0=0.7541; \ a=0.9346\}$ for sites and bonds respectively. Deviations
from available numerical
estimates are very small and range from $0$ to
$\pm 0.008$. At high dimensions a third class is found which recovers the
infinite Cayley
tree limit. It is defined by
$\{b=2a-1; \  p_0=2^{a-1}\}$ for both sites and bonds.
Using high dimensional hypercubic expansions [14, 17, 18] as data we found
$\{a=0.8800\}$ for
sites and $\{a=0.3685\}$ for bonds. The agreement is excellent with
deviations within $\pm 0.0004$.
The value of
dimension $d_c$
at which  the third class holds is between $d=6$ and $d=8$.
The results and perspectives are discussed.

From an analysis of the Bethe approximation on a regular lattice, we found
recently [13, 14]
an underlying lattice homogeneity breaking. On this basis we
obtained a new expression,
\begin{equation}
x_c=1/\sqrt{(q-1)}\:,
\end{equation}
for the percolation threshold, instead of the well known Bethe expression,
\begin{equation}
x_c^B=1/(q-1)\:.
\end{equation}

Eq. (1) yields results which are in good agreement with exact and numerical
site percolation thresholds at $d=2$.
However
results get poor at higher dimensions. This discrepancy evidences
a missing d-dependence which could be equal to one at $d=2$. It thus hints a
possible $(d-1)$ dependance. In parallel
Eq. (1) did single out a $(q-1)$ dependence. From the hypercube case we could
expect dimension
and coordinance to play a similar role. Therefore
one extension
of Eq. (1) is to
substitute the product
$(d-1)(q-1)$ for $(q-1)$.
On this basis, a natural generalization  of
Eq. (1) is the power law,
\begin{equation}
p_c=p_0[(d-1)(q-1)]^{-a}\:.
\end{equation}
It is worth to note that Eq. (3) reduces to Eq. (1) for the set $\{p_0=1; \
a=1/2 \}$ besides the variable multiplicator factor $(d-1)$ which
equals one at $d=2$.

At this stage, we report a log-log plot of known estimates $p_c$ [7, 19] as a
function of the associated product $(d-1)(q-1)$. The plot is shown in Fig.
(1) for the site case.
All the points align on only two straight lines, corresponding to the two
classes of lattices, above mentioned.

In parallel the log-log plot for bond data [7, 19] exhibits constant
deviations from
a straight line
at each dimension.
These deviations suggest
a dimension rescaling of the variable
$(d-1)(q-1)$ resulting in,
\begin{equation}
p_c=p_0[\frac {(d-1)(q-1)}{d}]^{-a}\:.
\end{equation}
The log-log plot using Eq. (4) is shown in Fig. (2). The situation is now
identical to
the site case, as all points align on two straight lines associated
to the two classes of lattices.
The percolation thresholds using Eqs. 3 and 4 are shown in Tables (1) and (2)
together with
exact results and numerical estimates
for respectively site and bond dilution, respectively.

At this stage we have a one exponent power law for sites (Eq. 3) and
bonds (Eq. 4). However
the site variable $(d-1)(q-1)$ has to be rescaled by dimension in the case
of bonds. The two expressions can thus be unified under the form
\begin{equation}
p_c=p_0[(d-1)(q-1)]^{-a}d\ ^b\:.
\end{equation}
 From Eqs (3) and (4) we have
$b=0$ for sites
and $b=a$ for bonds.
The agreement is remarkable and holds true from $d=2$ up to $d=7$
where threshold estimates are available.

To complete the discussion we compare our results to d-dimensional simple
hypercubic lattice
percolation thresholds derived as $1/(q-1)$ expansions.
For site percolation it is [16],
\begin{equation}
p_c^S=(q-1)^{-1}+\frac{3}{2}(q-1)^{-2}+\frac{15}{4}(q-1)^{-3}+\frac{83}{4}(q
-1)^{-4}+...\:,
\end{equation}
which becomes for bond percolation [17],
\begin{equation}
p_c^B=(q-1)^{-1}+\frac{5}{2}(q-1)^{-3}+\frac{15}{2}(q-1)^{-4}+57(q-1)^{-5}+.
..\:,
\end{equation}
In parallel from Eq. (5) we get the expansion,
\begin{equation}
p_c=p_0
2^{a-b}(q-1)^{b-2a}\{1+(a+b)(q-1)^{-1}+\frac{a^2+a+2ab+b^2-b}{2}(q-1)^{-2}+.
..\}\:.
\end{equation}

Eqs. (6) and (7) have the same first leading term $1/(q-1)$ which is the
Bethe result.
In our case we have,
\begin{equation}
p_0 2^{a-b}(q-1)^{b-2a}\ ,
\end{equation}
which becomes
$p_c=p_0 2^{a}(q-1)^{-2a}$
for sites with the exponent $2a=1.23$ and
$p_c=p_0 (q-1)^{-a}\:$
for bonds with $a=0.94$. Both cases are clearly different from the Bethe
result $1/(q-1)$.
The respective $1/(q-1)$ exponent is close to, but definitively
different from one. Therefore we do not recover the
Bethe asymptotic limit. Moreover the site-$p_c$ becomes smaller that the
bond-$p_c$ at
$d=11$
which is clearly non-physical.

On this basis, although our formula agrees perfectly with available
numerical data up to $d=7$
(see Tables 1 and 2),
a third class must exist at high dimensions. To
determine its
characteristics we first note that
the Cayley tree result $x_c^B=1/(q-1)$ is believed to be the exact
$d\rightarrow \infty$ asymptotic limit for both bonds and sites. It then
indicates that two
different constraints $b=0$ and $b=a$ for respectively sites and bonds
cannot hold at
high dimensions. The $1/(q-1)$ limit is recovered from Eq. (9) if, and only if
$b=2a-1$ together
with $p_0=2^{a-1}$. Eq. (5) becomes,
\begin{equation}
p_c=2^{a-1}[(d-1)(q-1)]^{-a}d\ ^{2a-1}\:,
\end{equation}
which gives a straight line in a log-log plot of $2dp_c$ versus
$2d^2/[(d-1)(q-1)]$.

We then determine the value of exponent $a$ using numerical estimates from
the $1/(q-1)$ expansions in Eqs. (6) and (7) which are supposed to be exact in the
$d\rightarrow \infty$ limit. 
From Fig. (3) we find
$a=0.8800$ for sites and $a=0.3685$ for bonds. It is worth noting that while 
Eqs. (6) and (7) are derived for hypercubes only, Eq. (10) holds for
any lattice at $d\geq d_c$.

From Fig. (3) the best estimate for the crossover dimension is $d_c=8$.
However we cannot preclude that $d_c$ is as small as
$d_c=6$ which is the upper critical dimension for percolation. 
Differences between numerical estimates and associated values of $p_c$ deduced, on the one hand from Eq. (3) 
or (4), and
on the other hand from  Eq. (10),
are comparable to numerical
errors (see Table (3) and Fig. (3)). From current data we have $6\leq d_c \leq 8$.
 Exact determination of $d_c$  
requires more accurate
numerical estimates of percolation thresholds at these dimensions.

Moreover the crossover at $d_c$ seems to be driven by 
a dimensional phenomena. This conjecture is based on the following observation.
At $d=5$, fcc lattice which has $(d-1)(q-1)=156$ belongs to the second class (see Fig. (1)).
In parallel hypercube at $d=9$ belongs to the third class (see Fig. (3)) though it has $(d-1)(q-1)=136$.

The various universality classes for
site and
bond dilution were found to be identical. One class is restricted to two dimensions and contains
triangle, square and honeycomb lattices. The second universality class
embodies 
 Kagom\'{e} and all lattices at $3\leq d\leq d_c$. It is interesting to stress that square and
Kagom\'{e} which both have $d=2$
and $q=4$ turn out to be in different classes. The third class embodies all lattices at $d\geq d_c$.
At $d=1$ ($q=2$) exact value $p_c=1$ for both sites and
bonds implies $a=0$.

At this stage it is worth to emphazise numerical values of
$p_0$ and $a$ have been determined using as input data what is denoted in Tables (1) and (2) 
as ``exact" thresholds.
Obviously these values will be different using a different set of input or
a restricted one.
We chose arbitrarily to use and report values form Refs. (6, 17).
We are here advocating the power law form of Eq. (5) rather than the third
decimal associated to
the determination of $p_0$ and $a$.

Error bars for numerical estimates are believed
to be on the last given figure. It should thus
not exceed $\pm 0.001$ in case $p_c$ is given with four decimals and $\pm 0.01$
in case it is given with three decimals. 
Therefore deviations $\Delta=p_c-p_c^e$  are significant at $d=2$ though they are small. 
For all lattices at this 
low dimension, $\mid \Delta \mid$ exceeds 0.001
($\Delta$ is only 0.0013 in the 
triangular lattice case, but there, $p_c$ is known exactly). 

To further investigate this anomaly at $d=2$,
we have explored percolation thresholds when the connection range includes
next (nnn) and next-next (nnnn) neighbors. Numerical estimates $p_c^e$ are
scarce
and restricted to site percolation [20]. They strongly suggest
all $d=2$-lattices enter the second universality class as soon as
the connection range exceeds nearest neighbors. Associated site percolation thresholds 
 are then calculated
using for $q$ the sum of sites within the connection range.   
We get respectively
$p_c$ = 0.388 for nnn-square, 0.294 for nnn-triangular, nnnn-square and
honeycomb, and 0.225 for nnnn triangle. It means $10^3\Delta = -1,\ +1,\  +2,
\  -6,\  0$  respectively, which is always smaller than the 
numerical estimate error bar $\pm 0.01$. However available square lattice site percolation
thresholds for longer ranges [21] don't belongs to this class. It shows that an additional
class should be added to account for long range interaction percolation (over nnnn).

At $d\geq 3$, errors $\Delta$ may not have a statistical significance.
Only the value of the bond percolation threshold for the simple
cubic lattice exceeds significantly 0.001. Moreover,
the random distribution
of deviations opposes the existence of a missing systematic correction to
Eq. (5).

Last but not least we found that our power law is also valid to predict Ising critical 
temperatures $T_c$. Indeed the phase transition nature of percolation makes
critical thresholds similar to critical temperatures.
Dealing with pair exchanges,
it is natural to use the bond percolation formula (Eq. 4) for the reduced
temperature $K_c=\frac{J}{k_BT_c}$, with J the exchange coupling.
 Numerical estimates for $K_c$ at $d=3$ are $K_c^e=0.2217,
\ 0.1575$, and $0.1021$ for sc, bcc and fcc lattices respectively [22].  The set
$\{p_0=0.6525;\ a=0.9251\}$ in Eq. (4) leads to critical temperatures
$K_c$ which depart from these values by the amount $K_c-K_c^e=+0.0012,\ +0.0008$,
and $-0.0006$, respectively. It is again comparable to the  
numerical estimate errors. 
%%%%%%%%%%%%%%%%%%%%%%%%%%%

\subsection*{Acknowledgments.}
We would like to thank Amnon Aharony and Dietrich Stauffer for
very stimulating comments and dicussions.

%%%%%%%%%%%%%%%%%%%%%%%%%%%%%%%%%%%%%%%%%%%%%%%%%%%%%%%%%%%%%%%%%%%%%%%
\newpage
{\LARGE References}\\ \\
1. {\sf S. R. Broadbent and J. M. Hammersley}, Proc. Camb. Phil. Soc.
\underline {53},
629 (1954)  \\
2. {\sf C. Domb}, Nature \underline {184}, 509 (1959) \\
3. {\sf P. J. Flory}, J. Am. Chem. Soc. \underline {63}, 3083, 3091, 3906
(1941)  \\
4. {\sf J. W. Essam}, Rept. Prog. Phys. \underline {43}, 833 (1980) \\
5. {\sf R . Zallen}, ``The Physics of Amorphous Solids", J. Wiley, New York
(1983)   \\
6. {\sf D. Stauffer and A. Aharony}, ``Introduction to Percolation Theory",
2nd Ed.,
Taylor and Francis, London (1994) \\
7. {\sf M. Sahimi}, ``Applications of Percolation Theory", Taylor and
Francis, London (1994)  \\
8. {\sf J. Adler}, Computers in Phys. \underline {8}, 287 (1994) \\
9. {\sf M. F. Sykes and J. W. Essam}, J. Math. Phys. \underline {5}, 1117
(1964) \\
10. {\sf M. E. Fisher and J. W. Essam}, J. Math. Phys. \underline {2}, 609
(1961) \\
11. {\sf M. Sahimi, B. D. Hughes, L. E. Scriven and T. Davis}, J. Phys.
A\underline {16}, L67
(1983)   \\
12. {\sf V. A. Vyssotsky, S. B. Gordon, H. L. Frish and J. M. Hammersley},
Phys. Rev.
\underline {123},1566 (1966)   \\
13. {\sf S. Galam and A. Mauger}, J. Appl. Phys. \underline {75/10}, 5526
(1994)   \\
14. {\sf S. Galam and A. Mauger}, Physica A\underline {205}, 502 (1994)   \\
15. {\sf R. E. Plotnick and R. H. gardner}, Lect. Math. in Life Sciences \underline {23}, 129 (1993) \\
16. {\sf D. S. Gaunt, M. F. Sykes and H. Ruskin}, J. Phys. A\underline {9},
1899 (1976)   \\
17. {\sf D. S. Gaunt and H. Ruskin}, J. Math. Phys. \underline {11}, 1369
(1978) \\
18. {\sf D. S. Gaunt and R. Brak}, J. Phys. A\underline {17}, 1761 (1984)   \\
19. {\sf D. Stauffer}, Physica A \underline {210}, 317 (1964) \\
20. {\sf J. W. Essam}, in ``Phase Transitions and Critical Phenomena "
V.\underline {2},
 Eds, C. Domb and M. S. Green, Academic Press, New York, 197 (1972) \\
21. {\sf M. Gouker and F. Family}, Phys. Phys. B\underline {28}, 1449
(1983)   \\
22. {\sf M. E. Fisher}, Rept. Prog. Phys. \underline {30}, 671 (1967)\\

%%%%%%%%%%%%%%%%%%%%%%%%%%%%%
%%%%%%%%%%%%%%%%%%%%%%%%%%%%%
\newpage
{\LARGE Figure captions}\\ \\

Figure 1: Inverse of site percolation thresholds
as a function of \\ $(q-1)(d-1)$ in logarithmic scales.\\ \\

Figure 2: Inverse of bond percolation thresholds
as a function of \\ $(q-1)(d-1)/d$ in logarithmic scales.\\ \\

Figure 3: Percolation thresholds in high-d hypercubes using both numerical
estimates for $d=6$ and $d=7$ and hypercubic expansions from
Refs. (14) and (15) at $d=8,\  9,...,\ 15,\  d=20,\  30,\  50$. The origin
corresponds
to the Cayley tree limit at infinite $d$. Straight lines are
according to Eq. (10). Broken curves correspond to Eqs. (3) and (4), for the
second universality class. \\ \\
\newpage
\begin{table}

\label{tbl}
\begin{center}
\begin{tabular}{|l|l|r|r|r|r|}
\hline
Dimension &Lattice &$\,q$&${p_c^e}$&${p_c}$&$\Delta$ \\ [5pt]
\hline
$d=2$ &Square & 4&  0.5928& 0.5984 & +0.0056\\ [5pt]
$\,$ &Honeycomb & 3&  0.6962& 0.6925  & -0.0037\\ [5pt]
$\,$ &Triangular & 6&  0.5000& 0.4979   & -0.0021\\ [5pt]
\hline
\hline
$d=2$ &Kagom$\acute{e}$*&  4& 0.6527 & 0.6540  & +0.0013\\ [5pt]
\hline
$d=3$ &Diamond& 4& 0.43& 0.43  & 0\\ [5pt]
$\,$ &sc& 6&  0.3116& 0.3115    &-0.0001\\ [5pt]
$\,$ &bcc& 8& 0.246& 0.253   & +0.007\\ [5pt]
$\,$ &fcc& 12 & 0.198& 0.192   & -0.006\\ [5pt]
\hline
$d=4$ &sc& 8& 0.197& 0.197  & 0\\ [5pt]
$\,$ &fcc*& 24 & 0.098& 0.095   & -0.003\\ [5pt]
\hline
$d=5$ &sc& 10& 0.141& 0.141& 0\\ [5pt]
$\,$ &fcc*& 40 & 0.054& 0.057   & +0.003\\ [5pt]
\hline
$d=6$ &sc& 12& 0.107& 0.109  & +0.002\\ [5pt]
$\,$ &fcc*& 60 &  & 0.039   &  \\ [5pt]
\hline
$d=7$ &sc& 14& 0.089& 0.088& -0.001\\ [5pt]
$\,$ &fcc*& 84 &  & 0.028   &  \\ [5pt]
\hline
\end{tabular}
\end{center}
\caption{\sf Site percolation thresholds from this work $p_c$ compared to
``exact estimates"
$p_c^e$ taken from [6, 17]. $\Delta \equiv p_c-p_c^e$. $\ast$ means not
included to determine
$p_0$ and $a$. The first universality class is defined by $p_0=0.8889$ and
$a=0.3601$.
The second universality class is defined by $p_0=1.2868$ and $a=0.6160$.}
\end{table}

%%%%%%%%%%%%%%%%%%%%%%%%%%%%%%%%%%%%%%%%%%%%%%%%%%%%%%%%%%%%%%%%%%%%%%%
\newpage
\begin{table}

\label{tbl}
\begin{center}
\begin{tabular}{|l|l|r|r|r|r|}
\hline
Dimension &Lattice &$\,q$&${p_c^e}$&${p_c}$&$\Delta$ \\ [5pt]
\hline
$d=2$ &Square & 4&  0.50000& 0.49581 & -0.00419\\ [5pt]
$\,$ &Honeycomb & 3&  0.6527& 0.6558  & +0.0031\\ [5pt]
$\,$ &Triangular & 6&  0.34729& 0.34859   & +0.00130\\ [5pt]
\hline
\hline
$d=2$ &Kagom$\acute{e}$*&  4& 0.5244 & 0.5162  & -0.0082\\ [5pt]
\hline
$d=3$ &Diamond& 4& 0.388& 0.394  & +0.006\\ [5pt]
$\,$ &sc& 6&  0.2488& 0.2448    & -0.0040\\ [5pt]
$\,$ &bcc& 8& 0.1803& 0.1787   & -0.0016\\ [5pt]
$\,$ &fcc& 12 & 0.119& 0.117   & -0.002\\ [5pt]
\hline
$d=4$ &sc& 8& 0.1601& 0.1601  & 0\\ [5pt]
$\,$ &fcc*& 24 & $\,$ & 0.0527 &$\,$\\ [5pt]
\hline
$d=5$ &sc& 10& 0.1182& 0.1192& +0.0010\\ [5pt]
$\,$ &fcc*& 40 & $\,$& 0.0303   & $\,$\\ [5pt]
\hline
$d=6$ &sc& 12& 0.0942& 0.0951  & +0.0009\\ [5pt]
$\,$ &fcc*& 60 & $\,$& 0.0198   & $\,$\\ [5pt]
\hline
$d=7$ &sc& 14& 0.07879& 0.07923& +0.00044\\ [5pt]
$\,$ &fcc*& 84 & $\,$& 0.0140   & $\,$\\ [5pt]
\hline
\end{tabular}
\end{center}
\caption{\sf Bond percolation thresholds from this work $p_c$ compared to
``exact estimates"
$p_c^e$ taken from [6, 17]. $\Delta \equiv p_c-p_c^e$. $\ast$ means not
included to determine
$p_0$ and $a$. The first universality class is defined by $p_0=0.6558$ and
$a=0.6897$.
The second universality class is defined by $p_0=0.7541$ and $a=0.9346$.}
\end{table}

%%%%%%%%%%%%%%%%%%%%%%%%%%%%%%%%
%%%%%%%%%%%%%%%%%%%%%%%%%%%%%%%%%%%%%%%%%%%%%%%%%%%%%%%%%%%%%%%%%%%%%%%
\newpage
\begin{table}

\label{tbl}
\begin{center}
\begin{tabular}{|l|l|r|r|r|r|}
\hline
Dimension  &$p_c$-site&$p_c^s$&$\frac{1}{q-1}$&$p_c^b$&$p_c$-bond \\ [5pt]
\hline
$d=6$ &0.1056& 0.1075& 0.0909& 0.0936  & 0.0920\\ [5pt]
\hline
$d=7$ &0.0873& 0.0882& 0.0769& 0.0785& 0.0777\\ [5pt]
\hline
$d=8$ &0.0744& 0.0748& 0.0667& 0.0676& 0.0672\\ [5pt]
\hline
$d=9$ &0.0648& 0.0650& 0.0588& 0.0595  & 0.0593\\ [5pt]
\hline
$d=10$ &0.0574& 0.0575& 0.0526& 0.0531& 0.0530\\ [5pt]
\hline
\end{tabular}
\end{center}
\caption{\sf High-d hypercube thresholds:  $p_c$-site and
$p_c$-bond from Eq. (10) and the ``hypercube" expansions
$p_c^s$ and $p_c^b$ from respectively [14] and [15]. At $d>10$
the difference is smaller than $10^{-4}$; even relative errors are negligible,
as it can be seen in Fig.3.}
\end{table}

%%%%%%%%%%%%%%%%%
\end{document}